\documentclass[conference]{IEEEtran}
\IEEEoverridecommandlockouts

\usepackage{cite}
\usepackage{amsmath,amssymb,amsfonts}
\usepackage{algorithmic}
\usepackage{graphicx}
\usepackage{textcomp}
\usepackage{tikz}
\usepackage{xcolor}
\usepackage[left=1.72cm,right=1.72cm,top=1.9cm,bottom=3cm]{geometry}
\usetikzlibrary{arrows.meta, positioning, calc}

\def\BibTeX{{\rm B\kern-.05em{\sc i\kern-.025em b}\kern-.08em
    T\kern-.1667em\lower.7ex\hbox{E}\kern-.125emX}}

\begin{document}

\title{RAISE: A Low-Frequency Space-Based Payload for Solar Radio and RFI Measurements on the SMiLE Mission}

\author{
\IEEEauthorblockN{
Sangeetha A\IEEEauthorrefmark{1},
Harsha Avinash Tanti\IEEEauthorrefmark{2},
Abhirup Datta\IEEEauthorrefmark{1},
Anshu Kumari\IEEEauthorrefmark{2},
Rajesh Bodade\IEEEauthorrefmark{3}
}
\IEEEauthorblockA{
\IEEEauthorrefmark{1}Indian Institute of Technology Indore, Indore, India 
}
\IEEEauthorblockA{
\IEEEauthorrefmark{2}Physical Research Laboratory, Udaipur Solar Observatory (USO), Udaipur, India
}
\IEEEauthorblockA{
\IEEEauthorrefmark{3}Military College of Telecommunication Engineering, Mhow, India \\
sangeetha250301@gmail.com, harshaavinashtanti@gmail.com, abhirup.datta@iiti.ac.in,\\ anshu@prl.res.in, and rajeshbodade@gmail.com
}
}

\maketitle

\begin{center}
\small\itshape
This is the accepted draft manuscript of the work presented at IEEE SPACE 2026. The final version of this article will be published in IEEE Xplore.
\end{center}

\begin{abstract}
We present the Radio-wave Apparatus for Investigating Solar \& Earth interference (RAISE),
a compact low-frequency radio payload proposed for hosting on the SMiLE mission.
Operating from Low Earth Orbit, RAISE targets a spectral regime that is largely inaccessible
from the ground due to ionospheric effects.
The payload is designed to enable space-based observations of low-frequency solar radio
emissions while simultaneously characterizing terrestrial and ionospheric radio frequency
interference in the near-Earth environment.
RAISE employs mode-dependent Earth-pointing and Sun-pointing observations to generate
dynamic spectral measurements relevant to space weather studies and low-frequency radio
mission planning.
As an experimental and technology demonstration payload, RAISE provides essential
heritage for future space-based low-frequency radio astronomy and space weather missions.
\end{abstract}

\section{Introduction}

Low-frequency radio observations provide direct insights into energetic processes in the
solar corona and its impact on near-Earth space.
Radio signatures of solar transients, called solar radio bursts, are intense non-thermal emissions generated by accelerated electrons
during solar flares and coronal mass ejections (CMEs) and are key diagnostics of particle
acceleration, shock formation, and magnetic reconnection and contain a plethora of information regarding. the local plasma conditions, such as temperature, density, and Mach number. The most common radio signatures of CMEs and flares are Type~II and Type~III solar radio bursts
In particular, the low-frequency components of 
Type~II and Type~III solar radio bursts, which
extend into the decametric and hectometric wavelength ranges, correspond to frequencies
below $\sim$30--50~MHz, where they trace shock-driven emission and fast electron beams
propagating through the outer corona and interplanetary medium\cite{dulk1985solar}.

Ground-based observations at these frequencies are severely limited by the Earth’s
ionosphere, which introduces absorption, refraction, and a sharp cutoff below
$\sim$10--30~MHz, depending on ionospheric conditions \cite{ansor2019ionosphere,burns2012lfreview}.
As a result, the lowest-frequency evolution of Type~II and Type~III solar radio bursts,
which carries critical information on CME-driven shocks and electron transport over large
coronal heights cannot be reliably accessed from the ground.
Space-based measurements are therefore essential for observing these burst components and
for obtaining an uninterrupted view of solar radio activity at low frequencies.

Early space missions such as the Radio Astronomy Explorer (RAE-1 and RAE-2) demonstrated
the scientific value of low-frequency radio observations from space, revealing solar and
planetary radio emissions inaccessible from Earth\cite{alexander1971rae1,alexander1975rae2}.
However, these measurements were conducted several decades ago under radio frequency
interference (RFI) conditions that differ significantly from the present-day near-Earth
environment.
Since then, the proliferation of terrestrial transmitters and spaceborne systems has
substantially altered the low-frequency radio landscape\cite{monstein2020rfi,offringa2010rfi}.

The Radio-wave Apparatus for Investigating Solar \& Earth interference (RAISE) is proposed
as a hosted payload on the SMiLE mission to address these limitations.
RAISE is a compact low-frequency radio instrument designed to operate in the
50~kHz--50~MHz frequency range from Low Earth Orbit (LEO).
Its primary scientific objective is to observe low-frequency solar radio bursts—especially
those inaccessible from the ground due to ionospheric effects, while simultaneously
characterizing terrestrial and ionospheric RFI from space.
In addition to its scientific role, RAISE serves as a technology demonstration payload,
providing critical instrumental and operational heritage for future low-frequency
space-based radio astronomy and space weather missions.

\section{Proposed Instrument}

The Radio-wave Apparatus for Investigating Solar \& Earth interference (RAISE) is a compact
low-frequency radio payload proposed for deployment on the SMiLE platform in Low Earth Orbit
(LEO).
RAISE is conceived as a self-contained radio observatory within a 3U form factor, designed
to operate with modest power requirements and high reliability under the constraints of a
hosted space mission.
The instrument is optimized to perform low-frequency solar radio observations and
systematic characterization of terrestrial and ionospheric radio frequency interference
(RFI) using a common, mode-dependent hardware architecture.

\subsection{Instrument Overview and Design Philosophy}

The design philosophy of RAISE emphasizes compactness, low power consumption, and operational
simplicity.
These considerations are driven by the need to operate below the ionospheric cutoff while
remaining compatible with the mass, volume, and power constraints of the SMiLE platform.
The payload supports multiple observation modes, including Earth-pointing and Sun-pointing
configurations, enabling targeted measurements of RFI and solar radio emissions without
requiring complex mechanical or electronic reconfiguration.

\subsection{Wideband Antenna System}

RAISE employs a compact, frequency-independent loop antenna designed to operate over the
50~kHz--50~MHz frequency range.
The antenna is intended to receive low-frequency solar radio emissions as well as terrestrial
and ionospheric radio signals, enabling continuous spectral coverage using a single antenna
system, consistent with established wideband antenna approaches in low-frequency radio
astronomy \cite{tanti2024biconical}.

To meet launch constraints, the antenna is stowed during launch and deployed after orbit
insertion using a low-shock motorized deployment mechanism.
In its deployed configuration, the antenna extends to less than 0.3~m, ensuring compatibility
with the SMiLE platform while maintaining sensitivity at low frequencies.
Electromagnetic compatibility is addressed through shielding and grounding strategies to
minimize spacecraft-generated interference and preserve a stable instrument noise floor in
the Low Earth Orbit environment.

\subsection{Receiver and Spectrometer}

Signals collected by the antenna are processed by a dedicated digital receiver and
spectrometer chain designed for low-frequency radio observations.
The spectrometer operates across the full 50~kHz--50~MHz band and provides a spectral
resolution of approximately 25~kHz, enabling time--frequency analysis of solar radio bursts
and RFI signatures.

The receiver architecture is designed to maintain sensitivity to weak signals while
accommodating strong in-band interference commonly encountered at low frequencies.
Filtering and dynamic range management are implemented to mitigate saturation during periods
of intense solar activity or elevated terrestrial RFI.

\subsection{Onboard Processing, Storage, and Telemetry}

RAISE incorporates an onboard digital signal processing (DSP) unit to manage the data rates
associated with continuous wideband observations.
The DSP performs real-time signal integration and lossless data compression, reducing raw
data volumes to levels compatible with the telemetry constraints of the SMiLE platform.

A solid-state storage system with a minimum capacity of 32~GB buffers science data during
periods without ground station contact.
Compressed data products are downlinked during scheduled passes, while telecommand support
enables payload configuration, mode switching, and calibration operations.

\begin{figure}[t]
    \centering
    \includegraphics[width=0.95\linewidth]{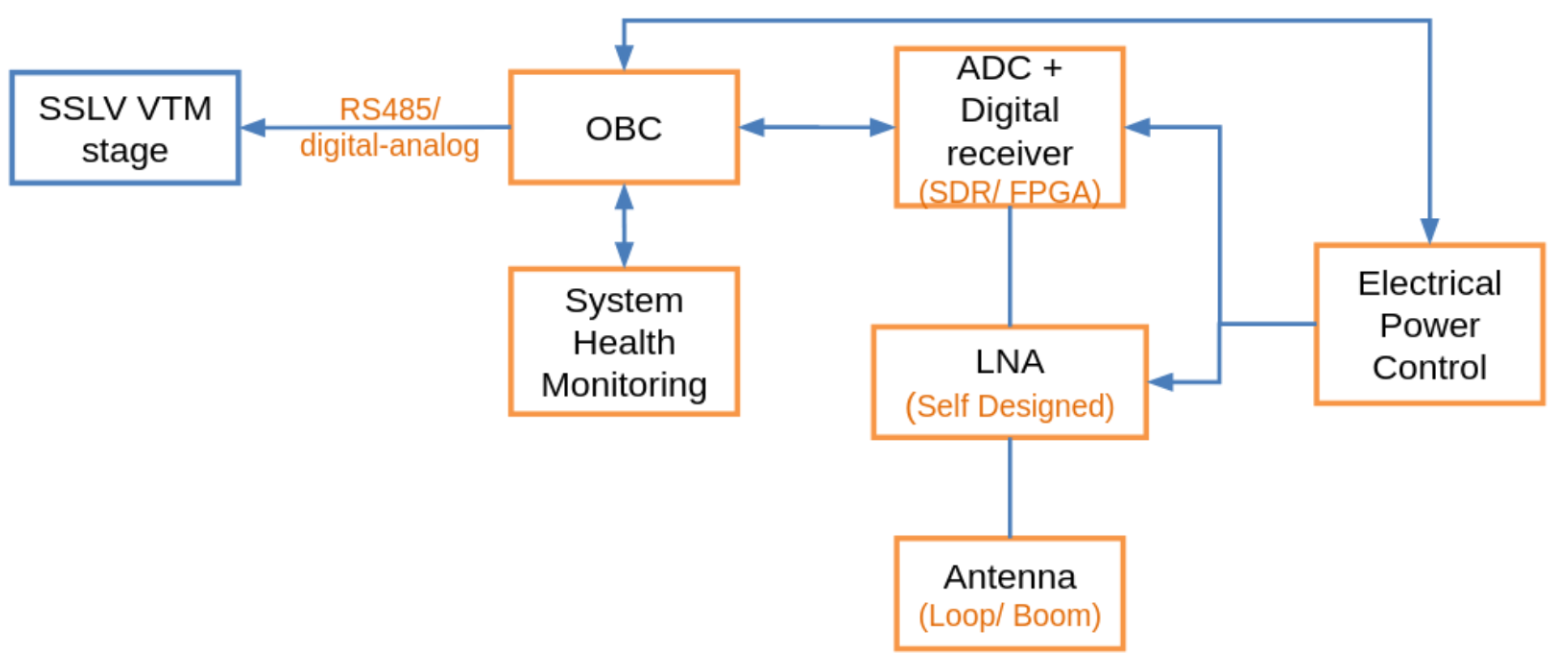}
    \caption{Functional block diagram of the RAISE payload showing the wideband antenna,
    analog front-end, digital receiver and spectrometer, onboard signal processing, data
    storage, and telemetry interface with the SMiLE platform.}
    \label{fig:raise_block}
\end{figure}

Figure~\ref{fig:raise_block} illustrates the functional architecture of the RAISE payload
and the signal flow from reception to onboard processing and telemetry.

\subsection{Heritage and Validation}

The RAISE instrument design builds on experience gained from the development and deployment
of compact sub-50~MHz ground-based radio instruments operated in the Arctic region
\cite{mapcon2025arctic}.
These deployments provided practical validation of wideband low-frequency antennas,
digital receiver architectures, onboard signal processing, and long-duration autonomous
operation in a challenging electromagnetic environment.

Complementary low-frequency RFI monitoring surveys conducted in remote and
interference-sensitive regions further demonstrated the stability and robustness of such
compact systems under realistic observing conditions \cite{ghosh2024arctic}.
Together, these efforts establish important heritage for space-borne low-frequency payloads.

Although ground-based experiments cannot replicate the space environment or remove
ionospheric effects, the experience gained from these Arctic deployments serves as a
pathfinder and risk-reduction step for RAISE.
The lessons learned directly inform antenna configuration, receiver robustness, and
operational strategies, providing confidence in the overall instrument architecture.

\subsection{Payload Specifications}

The RAISE payload has a total mass of approximately 10~kg, including the antenna, receiver
electronics, and deployment mechanism.
The payload conforms to a 3U volume envelope, with nominal power consumption of approximately
7~W during science operations and peak power requirements of up to 11~W during antenna
deployment and high-rate processing.

\section{RFI in Space and Solar Events}

Low-frequency radio observations from space open a unique window on both solar activity
and the radio environment surrounding Earth.
At frequencies below a few tens of megahertz, ground-based measurements are strongly
limited by ionospheric absorption and reflection.
As a result, access to this spectral regime relies almost entirely on space-based
instrumentation.

\subsection{Solar Radio Emissions Requiring Space-Based Observations}

The Sun is a strong and highly variable source of low-frequency radio emission.
During periods of enhanced solar activity, non-thermal processes associated with solar
flares and coronal mass ejections (CMEs) generate intense radio bursts that extend to
frequencies well below the terrestrial ionospheric cutoff.
These low-frequency components carry direct information about energetic electrons,
shock formation, and particle transport in the solar corona and interplanetary medium\cite{dulk1985solar}.

In particular, the low-frequency portions of Type~II and Type~III solar radio bursts
propagate to frequencies below $\sim$30~MHz, where ground-based observations are no longer
possible.
While higher-frequency components of some solar bursts can be detected from Earth,
their full spectral evolution can only be captured from space.
Space-based observations are therefore essential for studying the onset and development
of solar eruptive events and for improving space weather diagnostics\cite{reiner2001typeII,pulupa2010typeIII,burns2012lfreview,kasper2020sunrise}.

Table~\ref{tab:solar_bursts} summarizes representative solar radio emissions relevant to
the RAISE observing band and highlights the frequency ranges that require space-based
access.

\begin{table}[t]
\caption{Representative low-frequency solar radio emissions relevant to the RAISE observing band}
\centering
\begin{tabular}{|c|c|}
\hline
Emission Type & Typical Frequency Range \\
\hline
Type II (low-frequency component) & $\sim$1--30 MHz \\
Type III (low-frequency component) & $\sim$0.1--30 MHz \\
Type IV (low-frequency component) & $\sim$10--50 MHz \\
Low-frequency continuum emission & $<$30 MHz \\
\hline
\end{tabular}
\label{tab:solar_bursts}
\end{table}

Observations from Low Earth Orbit allow these emissions to be monitored without
ionospheric distortion, enabling continuous and broadband measurements that are not
feasible from the ground.

\subsection{Terrestrial and Ionospheric RFI in Space}

Alongside solar emissions, the low-frequency radio environment in near-Earth space is
strongly influenced by radio frequency interference originating from Earth and the
ionosphere.
Early space missions such as the Radio Astronomy Explorer (RAE-1 and RAE-2) provided the
first direct measurements of low-frequency radio noise above the ionosphere\cite{alexander1971rae1,alexander1975rae2}.
These missions established an initial picture of the radio environment in Earth orbit
and demonstrated the scientific potential of low-frequency radio observations from space.

However, the electromagnetic environment around Earth has evolved substantially since
those early surveys.
The growth of satellite constellations, global communication infrastructure, navigation
systems, and high-power terrestrial transmitters has led to a significant increase in
anthropogenic radio emissions.
As a result, existing low-frequency RFI maps derived from early space missions no longer
reflect present-day conditions\cite{monstein2020rfi,offringa2010rfi}.

RAISE is designed to address this gap by carrying out systematic and contemporary
measurements of the low-frequency RFI environment from Low Earth Orbit.
Operating in dedicated Earth-pointing modes, the payload will generate updated RFI
occupancy maps as a function of frequency, time, and orbital position.
These measurements will provide quantitative information on spectral occupancy and
temporal variability, which is essential for assessing the feasibility of future
space-based low-frequency radio missions.

By combining solar radio observations with an updated characterization of the near-Earth
radio environment, RAISE contributes both scientific measurements and practical knowledge
required for advancing low-frequency radio studies from space.

\section{Mission Details}

The RAISE payload is proposed as a hosted instrument on the SMiLE platform in Low Earth
Orbit (LEO) to enable low-frequency radio observations that are not feasible from the
ground.
At frequencies below a few tens of megahertz, the Earth’s ionosphere reflects and absorbs
incoming radio waves, effectively blocking access to a large fraction of the
low-frequency radio spectrum\cite{ansor2019ionosphere,burns2012lfreview}.
Operation from LEO places the payload above this ionospheric barrier, allowing direct
measurement of low-frequency solar radio emissions and characterization of the surrounding
radio frequency environment.

The preferred orbital configuration for RAISE corresponds to an altitude of
approximately 400--500~km with a near-equatorial inclination in the range of
$25^\circ$--$35^\circ$.
This orbit minimizes exposure to high-latitude ionospheric disturbances and auroral plasma
regions, leading to a more stable and repeatable radio environment.
Compared to ground-based observations, LEO also offers a cleaner electromagnetic setting
by reducing coupling with local terrestrial transmitters and site-specific interference,
while remaining operationally accessible for routine downlink and mission support.

Early space missions demonstrated the scientific potential of low-frequency radio
observations from above the ionosphere and provided the first measurements of the
near-Earth radio environment\cite{alexander1971rae1,alexander1975rae2}.
However, those surveys were carried out under radio conditions that differ significantly
from the present day\cite{monstein2020rfi}.
The rapid growth of satellite constellations, communication infrastructure, and
anthropogenic transmitters has fundamentally altered the low-frequency RFI environment.
At present, there is a lack of contemporary, space-based low-frequency RFI surveys,
particularly from Indian-led missions operating in this spectral regime.

RAISE is designed to address this gap by performing systematic measurements of the
low-frequency radio environment from LEO.
Through dedicated Earth-pointing observations, the payload will generate updated RFI
occupancy maps as a function of frequency, time, and orbital position.
These measurements will provide essential inputs for assessing spectral usability and for
informing the design and frequency planning of future space-based low-frequency radio
payloads.

Key mission-relevant parameters of the RAISE payload are summarized in
Table~\ref{tab:specs}, demonstrating compatibility with the SMiLE platform in terms of
mass, power, and operational requirements.

\begin{table}[t]
\caption{Key mission-relevant parameters of the RAISE payload}
\centering
\begin{tabular}{|l|c|}
\hline
Parameter & Value \\
\hline
Frequency Range & 50 kHz -- 50 MHz \\
Spectral Resolution & $\sim$25 kHz \\
Orbit Altitude & 400--500 km \\
Payload Mass & $\sim$10 kg \\
Nominal Power & $\sim$7 W \\
Peak Power & $\sim$11 W \\
Onboard Storage & $\geq$32 GB \\
Payload Volume & 3U \\
\hline
\end{tabular}
\label{tab:specs}
\vspace{-0.6cm}
\end{table}

\subsection{Operational Modes}

RAISE supports a small set of mode-dependent operations tailored to its mission objectives.
In Earth-pointing mode, the payload focuses on monitoring terrestrial and ionospheric radio
emissions to characterize the evolving low-frequency RFI environment.
In Sun-pointing mode, the payload is oriented toward the Sun to enable observations of
low-frequency solar radio activity from above the ionosphere.
Mode selection and scheduling are controlled through telecommands, allowing flexible
operation within the constraints of the SMiLE mission profile.

\subsection{Mission Phases}

Following orbit insertion, the mission begins with antenna deployment and initial system
health checks.
This is followed by in-orbit calibration to verify frequency response, gain stability, and
overall system performance.
Routine science operations commence after successful completion of these activities and
continue for the duration of the mission, supporting both targeted observations and
long-term monitoring of the low-frequency radio environment.

\section{Inference}

RAISE represents an experimental step toward establishing low-frequency radio observations
from space within the Indian space science program.
To date, there has been no Indian-led space mission operating dedicated instrumentation in
the sub-50~MHz regime, where access is fundamentally restricted by the terrestrial
ionosphere.
By targeting this largely unexplored spectral window from Low Earth Orbit, RAISE introduces
a new observational capability that complements existing ground-based and higher-frequency
space missions.

Although conceived as a compact and time-limited payload, the value of RAISE extends beyond
its operational lifetime.
As a technology demonstration, the mission validates the feasibility of compact, low-power
low-frequency radio instrumentation operating reliably in the near-Earth space environment.
The measurements obtained will establish a contemporary baseline of the low-frequency
radio-frequency interference environment, which is critical for the design, frequency
planning, and interference mitigation strategies of future space-based radio payloads\cite{burns2012lfreview,kasper2020sunrise}.

The outcomes of RAISE are expected to inform the development of next-generation space radio
astronomy and space weather missions by enabling improved mission planning and instrument
design at low frequencies.
In this sense, RAISE serves as a pathfinder mission, bridging current technological
capabilities and future low-frequency space observatories, and supporting sustained
exploration of the solar–terrestrial environment and the broader low-frequency radio
universe.

\bibliographystyle{IEEEtran}
\bibliography{refs}

\end{document}